\documentstyle[aps,epsfig]{revtex}

\begin{document}
\draft
\def\be{\begin{equation}}
\def\ee{\end{equation}}
\def\bfi{\begin{figure}}
\def\efi{\end{figure}}
\def\bea{\begin{eqnarray}}
\def\eea{\end{eqnarray}}
\def\vp{\varphi}
\title{Relaxation and Overlap Probability Function in the Spherical and Mean
Spherical Model}

\author{Nicola Fusco$^\dag$ and Marco Zannetti$^\S$}
\address{Istituto Nazionale per la Fisica della Materia,
Unit\`a di Salerno and Dipartimento di Fisica ``E.R.Caianiello'',
Universit\`a di Salerno,
84081 Baronissi (Salerno), Italy}

\maketitle

\dag nicola.fusco@sa.infn.it \S zannetti@na.infn.it

\begin{abstract}

The problem of the equivalence of the spherical and mean spherical models,
which has been thoroughly studied and understood in equilibrium, is
considered anew from the dynamical point of view during the time evolution
following a quench from above to
below the critical temperature. It is found that there exists a crossover
time $t^* \sim V^{2/d}$ such that for $t < t^*$ the two models are equivalent,
while for $t > t^*$ macroscopic discrepancies arise. The relation between
the off equilibrium response function and the structure of the equilibrium
state, which usually holds for phase ordering systems, is found to hold for
the spherical model but not for the mean spherical one. The latter model
offers an explicit example of a system which is not stochastically stable.

\end{abstract}

PACS: 05.70.Ln, 75.40.Gb, 05.40.-a

\section{Introduction} \label{sec1}

Let us consider a system with scalar continous order parameter
$\vp(\vec x)$ in a volume $V$ and hamiltonian
\be
{\cal H}[\vp(\vec x)]= \frac{1}{2}\int_V d \vec x
[(\nabla \vp)^2 + r\vp^2(\vec x)]
\label{1.1}
\ee
where $r \geq 0$. The spherical model of Berlin and Kac\cite{Berlin52}
is obtained by considering the extensive random variable
\be
\Psi= \frac{1}{V}\int_V d \vec x \vp^2(\vec x)
\label{1.2}
\ee
and computing equilibrium properties with the gaussian weight
$\rho_{\text{g}}[\vp]=\frac{1}{Z_{\text{g}}}e^{-\frac{1}{T}{\cal H}[\vp]}$
under the microcanonical constraint $\Psi = \alpha$,
where $\alpha$ is a given number. The mean spherical model of
Lewis and Wannier\cite{Lewis52}, instead, is obtained by imposing the constraint in the
mean, or canonically $\langle \Psi \rangle = \alpha$
which makes the model considerably easier to solve. The spherical model
was originally introduced by Berlin and Kac as an exactly soluble model
displaying critical phenomena. Subsequently, the enforcement of a
spherical constraint (in either form microcanonical or canonical) on the free
part of non linear problems has become an extremely useful and
practical way to generate mean field approximations.

However, despite
the great popularity of the method, it is usually overlooked that the
equilibrium
properties of the two models, as it was recognized early on, do coincide
above but not below the critical point. The origin of the discrepancy
was clarified first by Lax\cite{Lax55} and
further investigated by Yan and Wannier\cite{Yan65}.
Lastly, Kac and Thompson\cite{Kac77} showed how to
connect averages in the two models. As it is easy to understand,
the microcanonical and canonical constraint are equivalent as long as
the fluctuations
$\langle \delta \Psi^2 \rangle = \langle  \Psi^2 \rangle
-\langle \Psi \rangle^2$
are negligible. Indeed, above the critical temperature $T_C$ one has
the usual behavior for thermodynamic quantities
$\langle \delta \Psi^2 \rangle \sim 1/V.$
Not so below $T_C$, where fluctuations of $\Psi$ turn out to be
finite and independent of the volume
$\langle \delta \Psi^2 \rangle \sim \alpha^2.$

An important consequence of this, as we shall see in the following, is that
the nature of the mixed state below $T_C$ is quite different in the two models.
It is then interesting to investigate whether the two models are equivalent
or not when considering the time dependent properties in the relaxation process
following the quench from above to below the critical point. This is a
relevant question since, in practice, dynamics can be solved only in the mean
spherical case and in the literature it is given for granted that the
mean spherical form of dynamics applies to the the spherical
case as well.
On the basis of the previous considerations, clarification of this point
amounts to analyse the time evolution of $\langle \delta \Psi^2 \rangle$.
As we shall see, this depends on the order of the limits $t \rightarrow
\infty$ and $V \rightarrow \infty$. Namely, it turns out that if $V$ is
kept finite during
the relaxation there is a crossover from the preasymptotic behavior
$\langle \delta \Psi^2 \rangle \sim 1/V$ to the asymptotic one
$\langle \delta \Psi^2 \rangle \sim \alpha^2$ with the crossover time
$t^* \sim V^{2/d}$. Instead, if $V \rightarrow \infty$ from the outset,
then $\langle \delta \Psi^2 \rangle$ stays negligible for any finite time since
$t^*$ diverges. In any case, in the scaling regime $t < t^*$, the time
dependent evolution is identical in the two models, as usually assumed.

Although reassuring, this conclusion opens an unexpected and interesting
problem when it comes to test the connection between static and dynamic properties introduced by Cugliandolo and Kurchan\cite{Cugliandolo93}
for mean field spin glass and then
established in general by Franz, Mezard, Parisi and Peliti\cite{Franz98}
for slowly relaxing systems. Dynamic quantities are the autocorrelation
function $C(t,t_w)$ and the integrated linear response function
$\chi(t,t_w)$, while equilibrium properties are encoded into the overlap
probability function $P(q)$\cite{Mezard87}. The statement is that if
$\chi(t,t_w)$ depends on time through the autocorrelation function
$\chi(C(t,t_w))$, then one has
\be
\left . -T\frac{d^2 \chi(C)}{dC^2} \right )_{C=q} = \widetilde{P}(q)
\label{1.8}
\ee
where $\widetilde{P}(q)$ is the overlap probability function in the
equilibrium state obtained when the perturbation giving rise to
$\chi(t,t_w)$ is switched off. Therefore, the unperturbed overlap
probability $P(q)$ can be recovered from dynamics if $\widetilde{P}(q)$
coincides or it is simply related to $P(q)$.
This happens for stochastically stable systems\cite{Franz98}.
Formulated originally in the context of glassy systems, this
connection between statics and dynamics applies also to phase ordering
processes in non disordered systems\cite{Barrat98}.
A detailed account of the latter case
can be found in Ref.\cite{Corberi2001}. Now, the interesting point is that
the spherical and mean spherical model do share the same relaxation properties,
therefore the same $\chi(C)$ and the same $\widetilde{P}(q)$, while the
corresponding unperturbed overlap probabilities are
profoundly different. In particular, $P(q)$ can be recovered from
$\widetilde{P}(q)$ in the spherical model, but not in the mean spherical
one.

The paper is organized as follows. In Section 2 and in Section 3 the relevant
properties of the spherical and mean spherical models are presented.
The relation between the two models is discussed in Section 4 and comments
on the connection between statics and dynamics are made in the concluding
Section 5.

\section{Spherical Model} \label{sec2}

Dynamical evolution is described by the Langevin equation
\be
\frac{\partial \vp (\vec x,t)}{\partial t} =
-\frac{\delta{\cal H[\vp]}}{\delta \vp (\vec x,t)} + \eta (\vec x,t)
\label{2.1}
\ee
where $\eta (\vec x,t)$ is a gaussian white noise. Taking for
${\cal H[\vp]}$ the gaussian model~(\ref{1.1}) and
Fourier transforming with respect to space we have
\be
\frac{\partial \vp (\vec k,t)}{\partial t} =
-\omega_k \vp (\vec k,t) + \eta (\vec k,t)
\label{2.2}
\ee
with $\omega_k = k^2 + r$ and
\be
\begin{array}{ll}
\langle \eta (\vec k, t) \rangle = 0 \\
\langle \eta (\vec k,t) \eta (\vec k',t') \rangle
= 2T V\delta_{\vec k +\vec k',0} \delta (t-t')
                   \end{array}
 \label{2.4}
\ee
where $T$ is the temperature of the quench. Integrating~(\ref{2.2})
we obtain
\be
\vp (\vec k,t)= R_g(\vec k,t) \vp (\vec k,0)
+\int_0^t dt' R_g(\vec k,t-t') \eta (\vec k, t')
\label{2.5}
\ee
where $R_g(\vec k,t)= e^{-\omega_k t}$.
Considering an infinite temperature initial state with
\be
\begin{array}{ll}
\langle \vp (\vec k,0)  \rangle = 0 \\
\langle \vp(\vec k,0)\vp(\vec k',0)    \rangle
= \Delta V \delta_{\vec k +\vec k',0}
                   \end{array}
 \label{2.7}
\ee
and taking averages over the initial state and thermal noise, from~(\ref{2.5})
follows
$\langle \vp (\vec k,t)\rangle =0$,
$\langle \vp (\vec k,t) \vp(\vec k',t) \rangle =
C(\vec k,t) V \delta_{\vec k +\vec k',0}$
where the equal time structure factor is given by
\be
C(\vec k,t)=  R_g^2(\vec k,t) \Delta + \frac{T}{\omega_k}
[1- R_g^2(\vec k,t)].
\label{2.9}
\ee
With the initial condition~(\ref{2.7}) and the linear equation~(\ref{2.2})
the configuration $[\vp(\vec k,t)]$ executes a zero average gaussian
process whose probability distribution is given by
\be
\rho_g([\vp(\vec k)],t) = \frac{1}{Z_g(t)}e^{-\frac{1}{2V} \sum_{\vec k}
\vp(\vec k) C^{-1}(\vec k,t) \vp(-\vec k)}
\label{2.10}
\ee
where $Z_g(t)= \int d [\vp(\vec k)] e^{-\frac{1}{2V} \sum_{\vec k}
\vp(\vec k) C^{-1}(\vec k,t) \vp(\vec -k)}$.
From~(\ref{2.10}) one can compute the one time properties of the gaussian
model, including those at equilibrium obtained letting
$t \rightarrow \infty$.

Let us next define the joint probability of a configuration
$[\vp(\vec k)]$ and of a random variable $\Psi$ by
\be
\rho_g([\vp(\vec k)],\Psi,t)=\rho_g([\vp(\vec k)],t)
\delta \left ( \Psi - \frac{1}{V^2} \sum_{\vec k} \mid \vp(\vec k) \mid^2
\right ).
\label{2.12}
\ee
Clearly, $\rho_g([\vp(\vec k)],t)$ is recovered integrating over $\Psi$,
while the probability of $\Psi$ is given by
\be
\rho_g(\Psi,t) = \int d [\vp(\vec k)] \rho_g([\vp(\vec k)],\Psi,t).
\label{2.13}
\ee
Introducing the probability of $[\vp(\vec k)]$ conditioned to a given
value of $\Psi$
\be
\rho_g(\Psi \mid [\vp(\vec k)],t) = \frac{\rho_g([\vp(\vec k)],\Psi,t)}
{\rho_g(\Psi,t)}
\label{2.14}
\ee
we may also write
\be
\rho_g([\vp(\vec k)],t) = \int d \Psi \rho_g(\Psi,t)
\rho_g(\Psi \mid [\vp(\vec k)],t).
\label{2.15}
\ee
Notice that conditioning with respect to $\Psi$ is tantamount
to imposing the spherical constraint. Hence, the probabilty
distribution for the Berlin-Kac spherical model\cite{Berlin52} can
be written as
\be
\rho_s([\vp(\vec k)],t;\alpha) = \rho_g(\Psi=\alpha \mid [\vp(\vec k)],t).
\label{2.16}
\ee

\section{Mean Spherical Model} \label{sec3}

As stated in the Introduction, the mean spherical model is obtained by
imposing the constraint in the mean
$\frac{1}{V^2}\sum_{\vec k} \langle \mid \vp (\vec k) \mid^2 \rangle
= \alpha.$
This can be done by the Lagrange multiplier method or, what is the same,
by modifying~(\ref{2.2}) letting the parameter $r$ to be a function of
time
\be
\frac{\partial \vp (\vec k,t)}{\partial t} = - [k^2 + r(t)]\vp (\vec k,t)
+ \eta (\vec k,t)
\label{3.2}
\ee
where $r(t)$ is to be determined self consistently through the constraint,
which can be rewritten as
\be
\frac{1}{V}\sum_{\vec k} C(\vec k,t) = \alpha.
\label{3.3}
\ee
The structure of the solution of the equation of motion remains the same
\be
\vp (\vec k,t) = R_{\text{ms}}(\vec k,t,0)\vp(\vec k,0)
+\int_0^t R_{\text{ms}}(\vec k,t,t') \eta (\vec k,t')
\label{3.4}
\ee
where now
\be
R_{\text{ms}}(\vec k,t,t') = \frac{Y(t')}{Y(t)} e^{-k^2(t-t')}
\label{3.5}
\ee
with $Y(t) = e^{Q(t)}$ and $Q(t)=\int_0^t dt' r(t')$.
The equal time structure factor is given by
\be
C(\vec k,t) = R_{\text{ms}}^2(\vec k,t,0) \Delta +
2T \int_0^t dt' R_{\text{ms}}^2(\vec k,t,t')
\label{3.8}
\ee
where, clearly, the initial value $\Delta$ must also be consistent
with~(\ref{3.3}). This requires
$V^{-1}\sum_{\vec k} \Delta = \int \frac{d^d k}{(2\pi)^d} \exp (-k^2/\Lambda^2)
\Delta = \alpha$ where, as we shall always do in the following, in
transforming the sum over $\vec k$ into an integral we make explicit the
existence of an high momentum cutoff $\Lambda$. Hence, eventually
$\Delta = (4 \pi)^{d/2} \Lambda^{-d} \alpha$.
Finally, the probability distribution keeps the gaussian
form~(\ref{2.10})
\be
\rho_{\text{ms}}([\vp(\vec k)],t) = \frac{1}{Z_{\text{ms}}(t)}
e^{-\frac{1}{2V} \sum_{\vec k}
\vp(\vec k) C^{-1}(\vec k,t) \vp(-\vec k)}
\label{3.9}
\ee
where now $C(\vec k,t)$ is given by~(\ref{3.8}).

In order to have an explicit solution the function $Y(t)$ must be determined.
This is done in Appendix I where, for simplicity, the computation
has been limited to the case $2 <d < 4$. For large time one finds
\be
Y^2(t) = \left \{ \begin{array}{ll}
Be^{t/\tau}  & \mbox{ for $T>T_C$}  \\
B \left [e^{t/\tau} +(t/t^*)^{-\omega} \right ]    & \mbox{ for $T\leq T_C$}
                        \end{array}
               \right .
\label{3.10}
\ee
where
\be
T_C=\frac{2\Lambda^{d-2}}{(4\pi)^{d/2}(d-2)}\alpha
\label{3.11}
\ee
and the expressions of $B$,$\tau$,$\omega$ and $t^*$ are listed
in Appendix I. Here we point out that
$B$ and $\tau$ are independent of the volume for $T>T_C$, while
$B$ vanishes and $\tau$ diverges as $V \rightarrow \infty$  for $T \leq T_C$.
Inserting~(\ref{3.10})
in~(\ref{3.8}), for $T>T_C$ we find
\be
C(\vec k,t) = \frac{\Delta}{B}e^{-2(k^2+\frac{1}{2\tau})t} +
\frac{T}{k^2+\frac{1}{2\tau}}\left [ 1-e^{-2(k^2+\frac{1}{2\tau})t}
\right ]
\label{3.12}
\ee
and  for $T\leq T_C$
\be
C(\vec k,t) = \frac{\Delta}{B}\frac{(t/t^*)^{\omega}e^{-2k^2t}}
{\left [ 1+(t/t^*)^{\omega}e^{t/\tau} \right]} +
2T \int_{\hat{t}}^t dt' e^{-2k^2(t-t')}(t'/t)^{-\omega}
\left [ \frac{1+(t'/t^*)^{\omega}e^{t'/\tau}}
{1+(t/t^*)^{\omega}e^{t/\tau}} \right ]
\label{3.13}
\ee
where $\hat{t}$ is the microscopic
time necessary to elapse for~(\ref{3.10}) to apply. Notice that from Eq.~(\ref{6.18})
and from Eq.~(\ref{6.25}) follows that $t^*$ and $\tau$ are both $O(V^{2/d})$ for $T=T_C$,
while for $T<T_C$ and $d>2$ the two
time scales are separated with $t^* \sim V^{2/d} \ll \tau \sim V$.

In any
case, $\tau$ is the equilibration time. Taking $t \gg \tau$ from Eq.s~(\ref{3.12})
and~(\ref{3.13}) follows
\be
C(\vec k,t) = C_{eq}(\vec k) = \frac{T}{k^2+\xi^{-2}}
\label{3.14}
\ee
where the equilibrium correlation length $\xi$ is related to $\tau$ by
$\xi^2= 2\tau$. Hence, using Eq.~(\ref{6.18}) $T_C$ may be identified with the
static transition temperature separating the high temperature phase
where $\xi$ is independent of the volume, from the low temperature phase
where $\xi$ diverges with the volume
\be
\xi  \sim  \left \{ \begin{array}{ll}
    \left ( \frac{T-T_C}{T_C}\right )^{-1/(d-2)}
& \mbox{for $0<\frac{T-T_C}{T_C}\ll 1$ }      \\
        V^{1/d} & \mbox{for $T=T_C$}  \\
        V^{1/2} & \mbox{for $T<T_C$}.
                   \end{array}
               \right .
\label{3.15}
\ee

Finally, let us comment on the nature of the equilibrium state.
As Eq.~(\ref{3.9}) shows, each $\vec k$ mode is gaussianly distributed,
with zero average, at any time including at equilibrium.
Hence, in the low temperature phase the system does not order. The static
transition consists in the $\vec k =0$ mode developping a macroscopic
variance as the temperature is lowered from above to below
$T_C$\cite{Castellano97}
\be
\langle \vp_0^2 \rangle  \sim  \left \{ \begin{array}{ll}
    V  & \mbox{for $T > T_C$ }      \\
        V^{\frac{d+2}{d}} & \mbox{for $T=T_C$}  \\
        V^{2} & \mbox{for $T<T_C$}.
                   \end{array}
               \right .
\label{3.15bis}
\ee
For the sake of illustration let us consider $T=0$ where from
$C_{eq}(\vec k) = \alpha V \delta_{\vec k,0}$
follows
\be
\rho_{ms}[\vp (\vec k)] = \frac{1}{\sqrt{2\pi \alpha V^2}}
e^{-\frac{\vp_0^2}{2\alpha V^2}} \prod_{\vec k \neq 0}
\delta (\vp (\vec k)).
\label{3.18}
\ee

\section{The connection between the two Models} \label{sec4}

In order to explore how the spherical and the mean spherical model are
connected, let us rewrite Eq.~(\ref{3.9}) following
the same steps which have led to Eq.~(\ref{2.15}) and obtaining
\be
\rho_{\text{ms}}([\vp(\vec k)],t) = \int d \Psi \rho_{\text{ms}}(\Psi,t)
\rho_{\text{ms}}(\Psi \mid [\vp(\vec k)],t)
\label{4.1}
\ee
which yields the relation between the two models since the conditional
probability $\rho_{\text{ms}}(\Psi \mid [\vp(\vec k)],t)$ enforces the
constraint
{\it \`{a} la} Berlin and Kac. Actually, to be precise, this quantity is not
exactly the same as the spherical model distribution, since in Eq.~(\ref{2.16})
the constraint is imposed on the gaussian model, while here it is imposed
on the mean spherical model. In the following we will ignore the difference.

Looking at Eq.~(\ref{4.1}) the state in the mean spherical model can be
regarded as the mixture of states in the spherical model with constraint
values weighted by $\rho_{\text{ms}}(\Psi,t)$. The properties
of the two models are the same if this weight is narrowly peaked about the
mean value $\langle \Psi \rangle = \alpha$, while discrepancies are to be
expected if the weight spreads over significantly different values of
$\Psi$.
Therefore, the key quantity controlling the connection between the
two models is $\rho_{\text{ms}}(\Psi,t)$. The corresponding characteristic
function is
\be
\Theta (x) = \langle e^{ix\Psi} \rangle  =
e^{-\frac{1}{2} \sum_{\vec k} \ln [1-\frac{2}{V}ix C(\vec k,t)]}
\label{4.2}
\ee
where $\langle \cdot \rangle$ stands for the average computed
with~(\ref{3.9}).
Moments of $\Psi$ then are given by
$\left . \langle \Psi^n \rangle = \frac{d^n \Theta (x)}{d(ix)^n} \right
)_{x=0}$ and in particular
\be
\langle \Psi (t)\rangle = \frac{1}{V} \sum_{\vec k} C(\vec k,t) = \alpha
\label{4.3}
\ee
\be
\langle \delta \Psi^2 (t)\rangle = \langle \Psi^2 (t)\rangle - \langle \Psi (t)\rangle ^2 =
\frac{1}{V^2} \sum_{\vec k} C^2(\vec k,t).
\label{4.4}
\ee
Let us first consider what happens at equilibrium by letting $t \rightarrow \infty$
and inverting Eq.~(\ref{4.2}). Following Kac and Thompson \cite{Kac77}
and using Eq.s~(\ref{3.14}) and~(\ref{3.15}) for $T>T_C$ one finds
\be
\rho_{\text{ms}}(\Psi)=\frac{1}{\sqrt{2\pi \sigma}}
e^{-(\Psi-\alpha)^2/2\sigma}
\label{4.5}
\ee
with $\sigma = \frac{1}{V^2} \sum_{\vec k} C_{eq}^2(\vec k)$. For large volume this
can be rewritten as $\sigma = \frac{2}{V}\int \frac{d^dk}{(2 \pi)^d}
\frac{e^{-k^2/\Lambda^2}}{(k^2 + \xi^{-2})^2}$ and since the integral is finite
we have $\langle \delta \Psi^2 \rangle \sim O(1/V)$. Instead, for $T<T_C$ one finds
\be
\rho_{\text{ms}}(\Psi)     =  \left \{ \begin{array}{ll}
    0  & \mbox{for $\Psi < \alpha T/T_C $ }      \\
        \frac{e^{- \frac{(\Psi - \alpha T/T_C)}{2\alpha (1-T/T_C)}}}
        {\sqrt{2 \pi  (\Psi - \alpha T/T_C)\alpha (1-T/T_C)}}
        & \mbox{for $\Psi > \alpha T/T_C$}
                   \end{array}
               \right .
\label{4.6}
\ee
which gives
\be
\langle \delta \Psi^2 \rangle = 2[\alpha (1-T/T_C)]^2.
\label{4.7}
\ee

Hence, we have that in going from the high temperature to the low
temperature phase the fluctuations of $\Psi$ from microscopic become
macrocospic and, as anticipated above, significant differences must
be expected in the equilibrium states of the two models. In order
to see this explicitely, let us consider $T=0$ where the equilibrium state
in the mean spherical model is given by Eq.~(\ref{3.18}). The joint probability
of $[\vp (\vec k)]$ and $\Psi$ is given by
\be
\rho_{\text{ms}}([\vp(\vec k)],\Psi) = \frac{1}{\sqrt{2\pi \alpha V^2}}
e^{-\frac{\vp_0^2}{2\alpha V^2}} \prod_{\vec k \neq 0}
\delta (\vp (\vec k))
\delta \left (\Psi - \frac{1}{V} \int d \vec x \vp^2(\vec x)
\right)
\label{4.8}
\ee
which, due to the presence of the $\delta$ functions for $\vec k \neq 0$,
can be rewritten as
\be
\rho_{\text{ms}}([\vp(\vec k)],\Psi) = \frac{1}{\sqrt{2\pi \alpha \Psi}}
e^{-\frac{\Psi}{2\alpha }} \prod_{\vec k \neq 0}
\delta (\vp (\vec k))
\frac{V}{2 \sqrt{\Psi}} \left [
\delta(\vp_0 -V\sqrt{\Psi}) +\delta(\vp_0 +V\sqrt{\Psi}) \right ].
\label{4.11}
\ee
Identifying the first factor in the right hand side with the
$T \rightarrow 0$ limit
of Eq.~(\ref{4.6}), we find the conditional probability
\be
\rho_{\text{ms}}(\Psi \mid [\vp(\vec k)]) =\frac{1}{2} \left [
\delta(\vp_0 -V\sqrt{\Psi}) +\delta(\vp_0 +V\sqrt{\Psi}) \right ]
\prod_{\vec k \neq 0} \delta(\vp(\vec k))
\label{4.12}
\ee
which gives the $T=0$ state in the spherical model with a value $\Psi$
of the constraint. Comparing Eq.s~(\ref{3.18}) and~(\ref{4.12}), the difference
in the low temperature equilibrium states of the two models is quite
evident. While in the mean spherical model $\vp_0$ is gaussianly distributed,
in the spherical model the probability of $\vp_0$ is bimodal. The large
fluctuations~(\ref{4.7}) around the average of $\Psi$ do appear below $T_C$
since it is necessary to spread out the weight of $\Psi$ in order to
reconstruct a gaussian distribution by mixing bimodal distributions.

This difference in the structure of the low temperature equilibrium states
shows up quite clearly in the corresponding overlap probability
functions\cite{Mezard87}
\be
P(q)= \int d[\vp]d[\vp'] \rho[\vp]\rho[\vp']  \delta (Q[\vp,\vp'] - q)
\label{4.13}
\ee
where
\be
Q[\vp,\vp'] = \frac{1}{V} \int d \vec x \vp(\vec x)\vp'(\vec x)
= \frac{1}{V^2} \sum_{\vec k} \vp(\vec k)\vp'(-\vec k).
\label{4.14}
\ee
Keeping on considering, for simplicity, the $T=0$ states and looking
at the characteristic function
$\Theta(\lambda) = \int dq P(q) e^{i\lambda q}$ from Eq.s~(\ref{3.18})
and~(\ref{4.12}) with $\Psi=\alpha$ follows (Appendix II)
\be
\Theta_{\text{ms}}(\lambda) = \frac{1}{\sqrt{1+ (\lambda \alpha)^2}}
\label{4.15}
\ee
and
\be
\Theta_{\text{s}}(\lambda) = \cos (\lambda \alpha).
\label{4.16}
\ee
Inverting we have
\be
P_{\text{ms}}(q) = \frac{1}{\alpha \pi} K_0(\mid q \mid /\alpha)
\label{4.17}
\ee
where $K_0$ is the Bessel function of imaginary argument and
\be
P_{\text{s}}(q) = \frac{1}{2} [\delta (q-\alpha) + \delta (q+\alpha)].
\label{4.18}
\ee
The plots of these two functions in Fig.1 illustrate
the great difference in the ground states.

Having analysed the properties in the final equilibrium states, let us now
go back to the dynamical problem. Notice that at $t=0$, with $C(\vec k,0)
= \Delta$, from Eq.~(\ref{4.2}) follows that $\rho_{\text{ms}}(\Psi,t=0)$
is given by Eq.~(\ref{4.5}) with $\sigma=\frac{2\Delta^2\Lambda^d}
{(4\pi)^{d/2}V}$. Hence, in a quench to $T>T_C$ fluctuations of $\Psi$
remain microscopic throughout the time evolution, while in a quench to $T<T_C$
at some point $\langle \delta \Psi^2(t) \rangle$ must cross over from
$O(1/V)$ to $O(1)$. Again, for simplicity, we show this in the case of
the $T=0$ quench. From Eq.~(\ref{3.13}) in the limit $T \rightarrow 0$ we have
\be
C(\vec k,t) = \frac{\Delta (t/t_0)^{d/2} e^{-2k^2t}}
{[1+(t/t^*)^{d/2}]}
\label{4.19}
\ee
where $t_0=(2\Lambda^2)^{-1}$. Inserting into Eq.~(\ref{4.4})
we find for $t/t_0 \gg 1$ and $t^*/t_0 \gg 1$
\be
\langle \delta \Psi^2(t) \rangle = 2\alpha^2 (t/t^*)^d
\frac{[1+(t^*/2t)^{d/2}]}{[1+(t/t^*)^{d/2}]^2}
\label{4.20}
\ee
which gives
\be
\langle \delta \Psi^2(t) \rangle     =  \left \{ \begin{array}{ll}
     2\alpha^2 (t/2t^*)^{d/2}  & \mbox{for $t \ll t^* $ }      \\
         2\alpha^2 [1-(2-2^{-d/2})(t^*/t)^{d/2}]
        & \mbox{for $t \gg t^*$.}
                   \end{array}
               \right .
\label{4.21}
\ee
Therefore, $t^*$ is the crossover time separating the time regime $t \ll t^*$
with $\langle \delta \Psi^2(t) \rangle \sim O(1/V)$ from the time regime
$t \gg t^*$ with  $\langle \delta \Psi^2(t) \rangle \sim O((t^*)^{-d/2}) \sim O(1)$.

We may now comment on the non commutativity of the limits. If $V$ is kept
finite and the limit $t \rightarrow \infty$ is taken first, the system
equilibrates and the
two models at equilibrium are equivalent above $T_C$ but not
below. This conclusion keeps on remaining valid if, after having reached
equilibrium, the limit $V \rightarrow \infty$ is taken. If, instead, the limit
$V \rightarrow \infty$ is taken first then
$\langle \delta \Psi^2 \rangle \sim 1/V$ both for a
quench above and below $T_C$, since in the latter case the crossover
time $t^*$ diverges. Hence, the
two infinite volume models are always equivalent during relaxation.
Therefore, as stated in the Introduction, all dynamical quantities are the same
in the two models if the $V \rightarrow \infty$ limit is taken from
the outset.

\section{Concluding Remarks} \label{sec5}

We have analysed the relationship between the spherical and the
mean spherical model, at equilibrium and during relaxation from
an initial high temperature state to a lower temperature state. By
monitoring the behavior of the fluctuations of $\Psi$ we have
found that the two models are equivalent in a quench above $T_C$,
while discrepancies arise in a quench below $T_C$ if the volume
is kept finite and the time is larger than the crossover time $t^*
\sim V^{2/d}$. In the infinite volume limit $t^*$ diverges
and the relaxation dynamics is the same in the two models for all
times. In particular, the integrated auto-response function
\be
\chi(t,t_w) = \int_{t_w}^t dt' \int \frac{d^dk}{(2\pi)^d} R(\vec k,t,t')
e^{-k^2/\Lambda^2}
\label{5.1}
\ee
is expected to have the same form in both models. As anticipated in the
Introduction, this poses an interesting problem when considering the
connection between statics and dynamics.

In order to explain this it is necessary to expand somewhat on the behavior of
the autocorrelation and response function in a phase ordering
process\cite{Barrat98,Corberi2001,Bray94}. The basic feature is the split
of both these quantities into the sum of a stationary and an aging contribution
\be
C(t,t_w)= C_{\text{st}}(t-t_w) + C_{\text{ag}}(t/t_w)
\label{5.2}
\ee
\be
\chi(t,t_w)= \chi_{\text{st}}(t-t_w) + \chi_{\text{ag}}(t,t_w)
\label{5.3}
\ee
respectively due to thermal fluctuations and defect dynamics. The time
scales of these contributions are widely separated. In particular
$C_{\text{st}}(t-t_w)$ decays rapidly from $M_0^2 - M^2$ to zero
(where $M$ is the spontaneous magnetization at the temperature of the
quench and $M_0$ is the zero temperature spontaneous magnetization) while
$C_{\text{ag}}(t/t_w)$ decays from $M^2$ to zero on a much longer timescale.
The stationary terms in Eq.s~(\ref{5.2}) and~(\ref{5.3}) are related by the
equilibrium fluctuation dissipation theorem
\be
T\chi_{\text{st}}(t-t_w) = C_{\text{st}}(t,t) - C_{\text{st}}(t-t_w).
\label{5.4}
\ee
Rewriting the right hand side in terms of the full autocorrelation function
one finds\cite{Barrat98,Corberi2001}
\be
T\chi_{\text{st}}(t-t_w)     =  \left \{ \begin{array}{ll}
    M_0^2- C(t,t_w)  & \mbox{for $M^2 \leq C \leq M_0^2$ }      \\
        M_0^2 - M^2    & \mbox{for $C \leq M^2$}
                   \end{array}
               \right .
\label{5.5}
\ee
which yields
\be
\left . -T\frac{\partial^2\chi_{\text{st}}(C)}{\partial C^2} \right )_{C=q}=
\delta (q-M^2).
\label{5.6}
\ee

The next statement is that the aging contribution of the response function
obeys the scaling form
\be
\chi_{\text{ag}}(t,t_w) = t_w^{-a}\hat{\chi}(t/t_w)
\label{5.7}
\ee
and therefore vanishes for $t_w \rightarrow \infty$ if $a > 0$. As of yet
knowledge of the exponent $a$ remains limited. According to
euristic arguments\cite{Barrat98,Henkel2001} $a$ ought to coincide with the
exponent $\theta$ controlling the defect density $\rho(t) \sim
t^{-\theta}$, namely $\theta=1/2$ for scalar order parameter and
$\theta=1$ for
vector order parameter\cite{Bray94}, independently from dimensionality.
However,exact analytical results for the one dimensional Ising
model\cite{Godreche2000,Lippiello2000},
careful numerical computations\cite{Corberi2001} for the Ising model in
dimensions $d=2,3,4$ and exact analytical results for the large $N$
model\cite{Corberi2002} (which is equivalent to the mean spherical model)
give the non trivial behavior
\be
a  =  \left \{ \begin{array}{ll}
    \theta (\frac{d-d_L}{d_U - d_L}) & \mbox{for $ d < d_U$ }      \\
        \theta   & \mbox{for $d >d_U$}
                   \end{array}
               \right .
\label{5.11}
\ee
with logarithmic corrections at $d=d_U$. Here $d_L=1$ and $d_U=3$ for the
Ising model, while $d_L=2$ and $d_U=4$ in the large $N$ case. What happens
at $d=d_L$, where $a=0$, has been analysed in detail
in\cite{Lippiello2000,Corberi2001,Corberi2002}. Let us here consider
$d>d_L$, where $a>0$ and the aging contribution of the response function does
asymptotically vanish.  Then, putting together
Eq.s~(\ref{1.8}) and~(\ref{5.6}) one finds\cite{Corberi2002,Cugliandolo95}
\be
\widetilde{P}(q)= \delta (q-M^2).
\label{5.101}
\ee
Now, as stated in the Introduction,if stochastic stability holds
$\widetilde{P}(q)$ equals
the overlap probability function $P(q)$ of the unperturbed system, up to
the effects of global symmetries which might be removed by the perturbation.
This is what usually happens in a phase ordering system, like the Ising model,
where the perturbation brakes the up down symmetry and one has
\be
\widetilde{P}(q)= 2\theta(q) P(q).
\label{5.102}
\ee
Even if only half of the states are kept in the order parameter function
$\widetilde{P}(q)$ obtained from dynamics, using the symmetry $P(q)=P(-q)$
it is obviously possible to recover the full unperturbed overlap function.
Indeed, this is what takes place in the spherical model, where the
unperturbed overlap function~(\ref{4.18}) is given by the sum of two
$\delta$ functions and $\widetilde{P}(q)= 2\theta(q) P_{\text{s}}(q)$ holds.
Not so
in the mean spherical case, where the unperturbed overlap function~(\ref{4.17})
is non trivial and $\widetilde{P}(q) \neq 2\theta(q) P_{\text{ms}}(q)$
as it can
be seen at glance from Fig.1. Clearly, in the latter case
$P_{\text{ms}}(q)$ cannot
be reconstructed from knowledge of $\widetilde{P}(q)$. Therefore, stochastic
stability does not hold in the mean spherical model. It might be interesting
to investigate this point in other models treated with the spherical
constraint.

Acknowledgments - This work has been partially supported
from the European TMR Network-Fractals c.n. FMRXCT980183 and
from MURST through PRIN-2000.

\section{Appendix I}

In order to determine $Y(t)$
explicitely, let us rewrite Eq.~(\ref{3.3}) by separating out of the sum
the $\vec k =0$ term, as we expect it to become macroscopically
occupied at low temperature
\be
\frac{C(\vec k=0,t)}{V} + \frac{1}{V} \sum_{\vec k \neq 0} C(\vec k,t)
= \alpha.
\label{6.10}
\ee
For large volume the sum may
be approximated by an
integral and using Eq.s~(\ref{3.5}) and~(\ref{3.8}) we find
\be
\alpha Y^2(t) - \frac{2T}{V} \int_0^t dt' Y^2(t')
-2T \int_0^t dt' f(t-t' + \frac{1}{2\Lambda^2}) Y^2(t') =
\Delta \left [\frac{1}{V}+ f(t + \frac{1}{2\Lambda^2}) \right ]
\label{6.11}
\ee
where
\be
f(t + \frac{1}{2\Lambda^2}) = \int \frac{d^d k}{(2\pi)^d}
e^{-2k^2(t-\frac{1}{2\Lambda^2})}= \left [ 8\pi (t+\frac{1}{2\Lambda^2})
\right ]^{-d/2}.
\label{6.12}
\ee
The above equation can be solved
by Laplace transformation obtaining
\be
{\cal L}(z) = \frac{\Delta [1/V+ zh(z)]}{\alpha z - 2T[1/V + zh(z)]}
\label{6.13}
\ee
where ${\cal L}(z)$ and $h(z)$ are the Laplace transforms of $Y^2(t)$
and $f(t+\frac{1}{2\Lambda^2})$, respectively. The large time behavior
of $Y^2(t)$ is controlled by the small $z$ behavior of ${\cal L}(z)$.
For $2<d<4$ we have
\be
h(z)= K + \gamma z^{d/2-1} +O(z)
\label{6.14}
\ee
where $K= (4\pi)^{-d/2} \frac{\Lambda^{d-2}}{d-2}$ and $\gamma =
-(8 \pi)^{-d/2} \Gamma(1-d/2)$ are positive constants. Inserting in
Eq.~(\ref{6.13}) we have
\be
{\cal L}(z)=\frac{\Delta[1/V+Kz +\gamma z^{d/2}]}
{\alpha (1-T/T_C)z-2T\gamma z^{d/2} - 2T/V}
\label{6.15}
\ee
where $T_C=\frac{\alpha}{2K}=(4\pi)^{d/2}\frac{\alpha (d-2)}{2 \Lambda ^{d-2}}$.
Inverting the Laplace transform we have
\be
Y^2(t)= Be^{t/\tau} +I(t)
\label{6.16}
\ee
where the first contribution comes from the residue at the single pole
at $x_0=1/\tau$ on the positive real axis and
\be
I(t)= \frac{1}{2\pi i}\int_0^{\infty}dx [{\cal L}(x e^{-i\pi})
-{\cal L}(x e^{i\pi})]e^{-xt}
\label{6.17}
\ee
is the contribution from the cut along the negative real axis.
Looking for the zero of the denominator of Eq.~(\ref{6.13}) we
find
\be
\tau = \left \{ \begin{array}{ll}
    \left [ \frac{\alpha}{2T_C\gamma}(1-T/T_C)\right ]^{\frac{2}{2-d}}
& \mbox{  for $0<\frac{T-T_C}{T_C}\ll 1$ }      \\
        (\gamma V)^{2/d} & \mbox{  for $T=T_C$}  \\
       \frac{\alpha}{2T} (1-T/T_C)V
& \mbox{  for $T<T_C$}
                   \end{array}
               \right .
\label{6.18}
\ee
and computing the residue
\be
B=\Delta \alpha   \left \{ 2T\tau [ \alpha (d/2-1)(T/T_C-1)+dT\tau /V] \right \}^{-1}.
\label{6.19}
\ee
For $T>T_C$ the exponential dominates and the contribution from the cut
can be neglected in Eq.~(\ref{6.16}), since to leading order $B$ and $\tau$
are independent of the volume. Not so for $T \leq T_C$, where
taking into account the contribution from the cut we have
\be
Y^2(t) = B \left \{e^{t/\tau} + (t/t^*)^{-\omega}[1\pm (t/\tilde {t})^{\phi}]
\right \}
\label{6.20}
\ee
with
\be
\omega = \left \{ \begin{array}{ll}
    2-d/2 & \mbox{  for $T=T_C$ }      \\
        d/2 & \mbox{  for $T<T_C$}
                   \end{array}
               \right .
\label{6.21}
\ee
\be
\phi = \left \{ \begin{array}{ll}
    d/2 & \mbox{  for $T=T_C$ }      \\
        1 & \mbox{  for $T<T_C$}
                   \end{array}
               \right .
\label{6.22}
\ee
\be
B = \left \{ \begin{array}{ll}
    \frac{\Delta V^{1-4/d}}{2T_C^2\gamma^{4/d}}     & \mbox{  for $T=T_C$ }      \\
        \frac{\Delta}{\alpha V} (1-T/T_C)^{-2} & \mbox{  for $T<T_C$}
                   \end{array}
               \right .
\label{6.23}
\ee
\be
\tilde{t}= \left \{ \begin{array}{ll}
    \frac{1}{8\pi} \left [\frac{(d/2-1) \Gamma^2 (2-d/2) V}{2 \cos (d\pi/2) \Gamma (2-d/2)} \right ]^{2/d}
          & \mbox{, for $T=T_C$} \\
        \left [\frac{ \alpha (1-T/T_C)\Gamma (d/2)}{4T \Gamma (d/2-1)V} \right] & \mbox{, for $T<T_C$}
                   \end{array}
               \right .
\label{6.24}
\ee
\be
t^*=  \left \{ \begin{array}{ll}
       \left [(d-2)^2 \Delta ^{-2} (8\pi)^{d-2} \gamma ^{8/d} \right ]^{1/(d-4)}
             V^{2/d}
                & \mbox{ for $T=T_C$} \\
            \frac{V^{2/d}}{8\pi}  & \mbox{ for $T<T_C$}
            \end{array}
               \right .
\label{6.25}
\ee
and where in Eq.~(\ref{6.20}) the $+$  and $-$ sign apply to $T=T_C$ and
$T<T_C$, respectively.
Notice that in all cases $t^* \leq \tau$ and $\tilde{t} \geq t^*$,
therefore the dominant contribution is given by
\be
Y^2(t) \sim \left \{ \begin{array}{ll}
    t^{-\omega} & \mbox{ for $t < t^*$ }      \\
        e^{t/\tau} & \mbox{ for $t > t^*$}.
                   \end{array}
               \right .
\label{6.26}
\ee

\section{Appendix II}

Inserting Eq.~(\ref{3.18}) into the definition (\ref{4.13}) we obtain
for the overlap function in the mean spherical model at $T=0$
\begin{equation}
  P_{\text{ms}}(q)= \frac{1}{2\pi \alpha V^2} \int d\vp_0 d\vp'_0 e^{-\frac{\vp_0^2+\vp_0'^2}
  {2\alpha V^2}} \delta (\frac{\vp_0\vp'_0}{V^2} - q).
\label{7.1}
\end{equation}
The corresponding characteristic function is given by
\begin{equation}
  \Theta_{\text{ms}}(\lambda) =
    \frac{1}{2\pi \alpha V^2}  \int d\vp_0 d\vp'_0 \exp\{-\frac{1}
  {2\alpha V^2}(\vp_0^2+\vp_0'^2-2i\alpha\lambda\vp_0\vp'_0)\}
\label{7.2}
\end{equation}
and going over to polar integration variables one finds
\begin{eqnarray}
  \Theta_{\text{ms}}(\lambda) & = &
  \frac{1}{2\pi \alpha V^2} \int_{0}^{2\pi} d\vartheta \int_{0}^{\infty} dr r
\exp\{-\frac{r^2}
  {2\alpha V^2}(1-2i\alpha\lambda\sin\vartheta\cos\vartheta)\} \nonumber \\
& =  & \frac{1}{\pi} \int_{0}^{2\pi}\frac{d\vartheta}{2-\alpha\lambda e^{i\vartheta}+\alpha\lambda e^{-i\vartheta}}.
\label{7.3}
\end{eqnarray}
This integral can be rewritten in the form
$\frac{1}{i\pi}\oint_{\gamma}\frac{dz}{2z-\alpha\lambda
z^2+\alpha\lambda}$ where $\gamma$ is the circle of radius one
with center in the origin of the complex plane.  Since there is
a simple pole at $z_{0}=\frac{1-\sqrt{1+\alpha^2\lambda^2}}{\alpha\lambda}$
inside $\gamma$ we obtain
\begin{equation}
  \Theta_{\text{ms}}(\lambda) = \frac{1}{\sqrt{1+ (\lambda\alpha)^2}}
\label{7.4}
\end{equation}
and inverting the Fourier transform we find
\begin{equation}
  P_{\text{ms}}(q)
  =\frac{1}{2\pi\alpha}\int_{-\infty}^{\infty}dx\frac{e^{-i\frac{q}{\alpha}x}}{\sqrt{1+x^2}}.
\label{7.4}
\end{equation}
Taking into account that there are branch points at $\pm i$ and that the integration
contour is closed in the negative imaginary half plane for $q >0$ and viceversa
for $q < 0$, eventually we obtain
\begin{equation}
  P_{\text{ms}}(q) =\frac{1}{\pi\alpha}\int_{1}^{\infty}dy\frac{e^{-\frac{\mid q \mid}
{\alpha}y}}{\sqrt{y^2-1}}=
  \frac{1}{\pi\alpha}K_{0}(\frac{\mid q \mid}{\alpha})
\end{equation}

\noindent where $K_{0}$ is the Bessel function of imaginary
argument.

\clearpage

\begin{figure}
\begin{center}
\includegraphics[height=12cm]{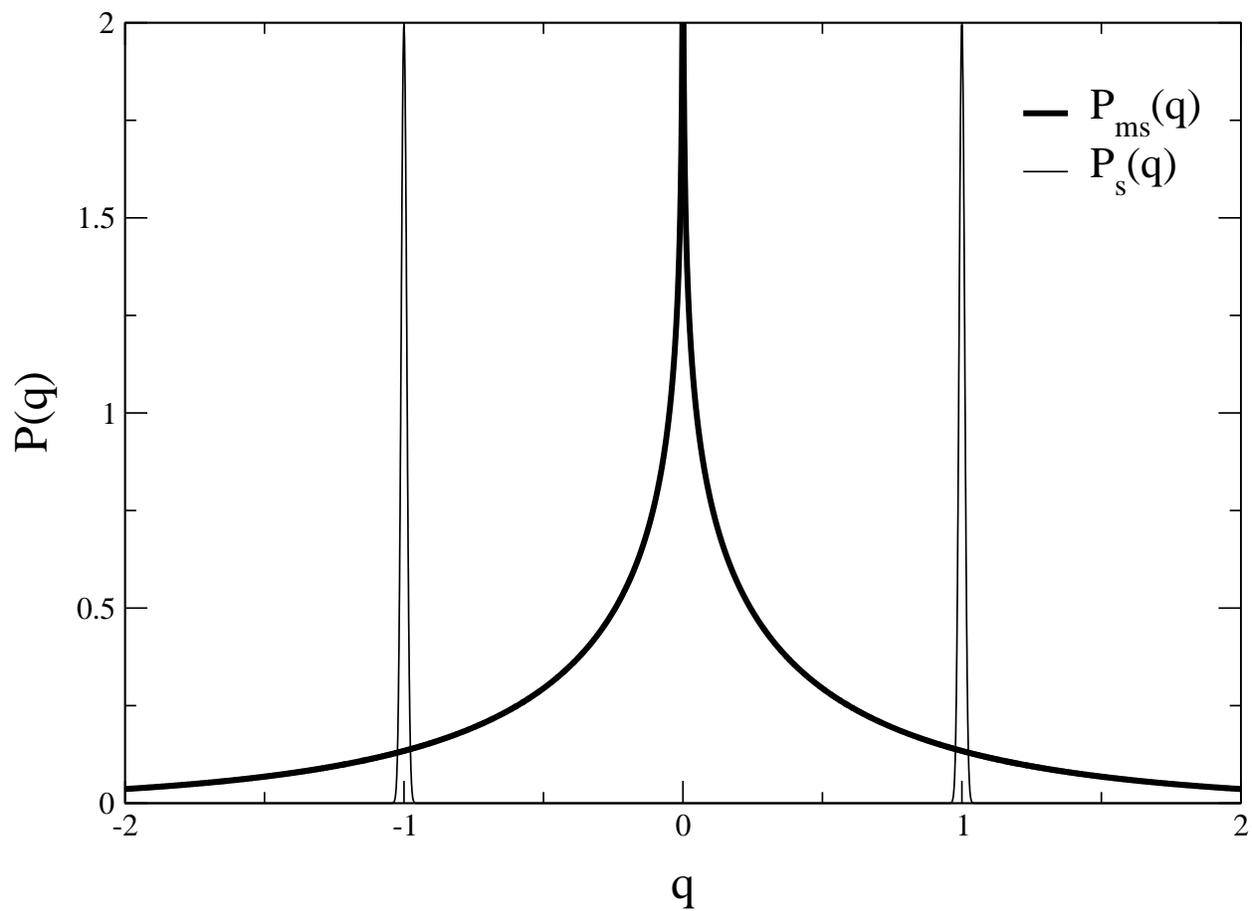}
\caption{The $T=0$ overlap probability function for the spherical and
the mean spherical model with $\alpha =1$.}
\end{center}
\end{figure}

\end{document}